\documentclass[conference]{IEEEtran}
\IEEEoverridecommandlockouts
\usepackage{cite}
\usepackage{amsmath,amssymb,amsfonts}
\usepackage{algorithmic}
\usepackage{graphicx}
\usepackage{textcomp}
\usepackage{xcolor}
\def\BibTeX{{\rm B\kern-.05em{\sc i\kern-.025em b}\kern-.08em
    T\kern-.1667em\lower.7ex\hbox{E}\kern-.125emX}}

\usepackage{hyperref}
\hypersetup{
    colorlinks = true, 
    urlcolor = black, 
    linkcolor = blue, 
    citecolor = blue 
}
\newcommand\numberthis{\addtocounter{equation}{1}\tag{\theequation}}

\begin{document}

\title{Compilation Techniques for Spin Qubits in a Shuttling Bus Architecture
\thanks{This work was supported by the European Commission (QUADRATURE: HORIZON-EIC-2022-PATHFINDEROPEN-01-101099697, WINC: HORIZON-ERC-2021-101042080), the Spanish Ministry of Science, Innovation and Universities (Beatriz Galindo program 2020, BG20-00023) and European ERDF (PID2021-123627OB-C5, QCOMM-CAT-Planes Complementarios), funding from MICIN and NextGenerationEU (PRTR-C17.I1), Generalitat de Catalunya, the ICREA Academia Award 2024, Science Foundation Ireland under Grant 21/RP-2TF/10019, and the predoctoral FPI-UPC grant supported by UPC and Banco Santander.}
}

\author{
    \IEEEauthorblockN{
        Pau Escofet\IEEEauthorrefmark{1},\
        Andrii Semenov\IEEEauthorrefmark{2},\
        Niall Murphy\IEEEauthorrefmark{2},\
        Elena Blokhina\IEEEauthorrefmark{2}\IEEEauthorrefmark{3},\\
        Sergi Abadal\IEEEauthorrefmark{1},\
        Eduard Alarcón\IEEEauthorrefmark{1},\ and
        Carmen G. Almudéver\IEEEauthorrefmark{4}
    }
    \IEEEauthorblockA{
        \textit{\IEEEauthorrefmark{1}Universitat Politècnica de Catalunya, Spain} \quad
        \textit{\IEEEauthorrefmark{2}Equal1 Labs, Ireland} \quad \\
        \textit{\IEEEauthorrefmark{3}University College Dublin, Ireland} \quad
        \textit{\IEEEauthorrefmark{4}Universitat Politècnica de València, Spain}\\
        pau.escofet@upc.edu
        \vspace{-10pt}
    }
}

\maketitle

\begin{abstract}
In this work, we explore and propose several quantum circuit mapping strategies to optimize qubit shuttling in scalable quantum computing architectures based on silicon spin qubits. Our goal is to minimize phase errors introduced during shuttling operations while reducing the overall execution time of quantum circuits. We propose and evaluate five mapping algorithms using benchmarks from quantum algorithms. The Swap Return strategy emerged as the most robust solution, offering a superior balance between execution time and error minimization by considering future qubit interactions. Additionally, we assess the importance of initial qubit placement, demonstrating that an informed placement strategy can significantly enhance the performance of dynamic mapping approaches.
\end{abstract}


\section{Introduction}
Quantum computing \cite{nielsen_chuang_2010} offers a new paradigm for processing information. Unlike classical bits, qubits can exist in multiple states simultaneously, enabling quantum algorithms to solve certain problems that are intractable for classical computers \cite{shor_polynomial_1997, Santagati2024}. As the field advances, efforts to implement scalable quantum computers have focused on various qubit technologies \cite{Nakamura_1999, kok_linear_2007, imagog_quantum_1999, cirac_quantum_1995}, with silicon spin qubits \cite{wild2012few} being particularly promising due to their compatibility with existing semiconductor manufacturing and fast operation times.

Quantum circuits must be compiled for physical hardware before execution, which involves native gate-set decomposition, %
initial qubit placement, and routing operations to ensure two-qubit interactions are satisfied.

Compilation strategies are typically tailored to specific architectures, leveraging their unique communication primitives and optimizing relevant performance metrics. For instance, heuristic methods have been developed for large-scale superconducting systems \cite{li2019tackling} or ion trap architectures \cite{schoenberger2024shuttlingscalabletrappedionquantum}. Other approaches focus on mapping quantum error correction codes to constrained architectures \cite{Helsen_2018} or addressing compilation challenges in distributed quantum architectures \cite{escofet2024revisiting}. Compilers for spin qubits are beginning to emerge, typically tailored to specific architectures \cite{Crawford2023, 10.1145/3624484}.

This work focuses on strategies to optimize qubit shuttling in silicon spin qubit architectures based on a shuttling bus, specifically we model a conveyor belt~\cite{Seidler2022} system. To address this, we propose several mapping strategies designed to tackle the qubit shuttling problem in a scalable manner. We also introduce an initial placement mapping based on the circuit's structure. Through benchmarking of quantum algorithms, we show that the \emph{Swap Return} strategy provides the best overall performance, striking an optimal balance between speed and error minimization, thus offering a robust solution for practical quantum circuit execution.

\section{Background}
\subsection{Silicon Spin Qubits}
Silicon spin qubits \cite{wild2012few} store information in the spin state of a single electron confined in a silicon quantum dot. The spin can be in a "spin-up" or "spin-down" state, corresponding to the $|0\rangle$ and $|1\rangle$ states. These qubits are controlled using magnetic or electric fields to manipulate the electron's spin, allowing for single- and two-qubit operations \cite{yoneda2018quantum}. Silicon-based spin qubits are typically created in silicon-germanium (Si/SiGe) heterostructures or silicon metal-oxide-semiconductor (MOS) devices, where precise control over the electron's environment ensures long coherence times and high-fidelity operations \cite{yoneda2018quantum}. Recent advancements have demonstrated that single- and two-qubit gate fidelities in silicon spin qubits have already exceeded the threshold for quantum error correction \cite{xue2022quantum}.

Silicon spin qubits are a promising qubit technology due to their compatibility with existing semiconductor manufacturing processes, which allows for easier integration with classical control electronics \cite{zwanenburg2013silicon}. This industrial compatibility, combined with silicon's excellent material properties~\cite{neumann2015simulation}, makes silicon spin qubits ideal for scaling up quantum processors.

Several techniques for qubit shuttling have been proposed for silicon spin qubits, including the bucket brigade (BB) method \cite{Mills2019}, the use of surface acoustic waves \cite{jadot2021distant}, and the conveyor belt approach \cite{Seidler2022}. Each method offers different advantages in terms of scalability, coherence preservation, and operational speed, with the conveyor belt method emerging as a promising solution for scalable architectures.

\subsection{Shuttling Bus Architecture with the Conveyor Belt Method}

Various types of shuttling bus architectures for spin qubits have been proposed in the literature~\cite{taylor2005fault,hollenberg2006two,Seidler2022}. This type of architecture suggests designated locations for single and two-qubit gates and a spin bus (also called a shuttling bus) for coherent transfer of a qubit state. In this work, we study an architecture based on the concept of the shuttling bus, specifically with the conveyor belt shuttling method. The operation of the shuttling bus and the fidelity of the state transfer are key aspects of this architecture. In the BB mode, electron qubits are transferred through tunnel-coupled quantum dots, which requires fine-tuning each dot pair and corresponding control lines, making it impractical for large arrays due to increased complexity and sensitivity to disorder \cite{langrock2023blueprint, yoneda2018quantum}. In the case of acoustic waves, the shuttling velocity is fixed, limiting the operation's flexibility, and the spin qubit decoherence is very hard to reduce \cite{bluhm2011dephasing}. In contrast, the conveyor belt uses a smooth electrostatic potential to shuttle qubits along a 1D channel, reducing the need for fine-tuning and minimizing the number of control signals, making it better suited for large-scale quantum computing \cite{langrock2023blueprint}.

In \cite{langrock2023blueprint}, Langrock et al. propose an error model for the conveyor belt shuttling process that analyzes key sources of decoherence, such as non-adiabatic transitions, spin dephasing, and orbital or valley excitations. The model examines how shuttling speed affects exposure to noise and disorder, with faster speeds reducing dephasing but increasing the risk of excitations to higher energy states. 

\begin{figure}
    \centering
    \includegraphics[width=\linewidth]{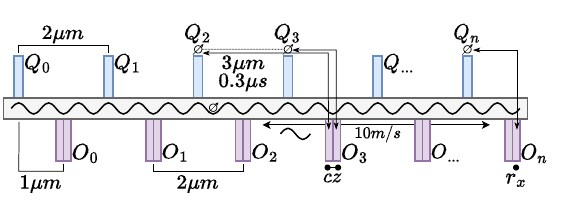}
    \vspace{-26pt}
    \caption{Sketch of the proposed shuttling belt architecture. Qubits $Q_2$ and $Q_3$ are being shuttled to manipulation zone $O_3$ at 10 m/s for a two-qubit gate, and $Q_n$ is being shuttled to $O_n$. Physical qubits are 2 µm apart, and manipulation zones are between every two qubits; thus, $Q_2$ is being shuttled 3µm, taking 0.3µs.}
    \vspace{-13pt}
    \label{fig:architecture}
\end{figure}

The proposed quantum computing architecture leverages a one-dimensional (1D) array of silicon spin qubits connected through a shuttling bus mechanism to allow for scalable and coherent qubit operations. The architecture is inspired by~\cite{taylor2005fault}. To enhance the architecture, we add locations for qubit storage and use the conveyor belt as a specific example of the shuttling bus. In this design, qubits are stored as the electron's spin state, which is transported along the array via a potential minimum that moves smoothly in a controlled manner, also referred to as the conveyor belt mode. Between the qubits, designated spaces facilitate single- and two-qubit gates and readout operations. A sketch of such architecture is depicted in Figure \ref{fig:architecture}, where qubits are shown in blue and manipulation zones in purple.

We use the phase error model proposed in \cite{langrock2023blueprint} to assess the dephasing that shuttling adds to the qubit, a metric to be minimized in the compilation process. The model for estimating the behaviour of the phase error $\delta C$ of a qubit moving with velocity $v$ for distance $L_s$ in a quantum dot array with linear dot size $L_{dot}$ is:

\vspace{-12pt}
\begin{align*}
    \delta C \sim& 2 \frac{l_c^{\delta \omega} L_s}{(v T_2^*)^2} + \frac{10^{-4}}{v} + 0.01  \left( \frac{1}{2} \frac{(\hbar a_x v)^2}{E_{vs,0}^2} e^{\frac{(a_x L_{dot})^2}{2}} \right) \\
    & + 0.01 \left( \frac{L_s}{\bar{d}} e^{-0.03 \ln 10\frac{E_{vs,0} L_{dot}}{\hbar v}} \right), \numberthis 
\label{eq:error_model}
\end{align*}
where the first term takes into account g-factor fluctuations (due to Overhauser fields and/or low-frequency $1/f$ charge noise) with correlation length $l_c^{\delta\omega}$ and static $T_2^*$ time; the second term is an upper bound for the effect of adiabatic passage hotspot through the Landau-Zener spin-valley anti-crossing using typical values for spin-orbit coupling due to magnetic field gradient; the third and the fourth terms take into account valley relaxation effects due to electron-phonon coupling in the presence of high and low density of atomistic irregularities, respectively, with average separation $\bar{d}$, which lead to a valley phase gradient $a_x$ along the $x$ axis ($E_{vs,0}$ is a valley splitting in the absence of the gradient).

\section{Compilation Techniques}

In this architecture, the compilation process begins by decomposing the input quantum circuit into the basis gates of the system, which we assume to be (\texttt{rx}, \texttt{rz}, \texttt{h}, \texttt{cz}) \cite{Weinstein2023, amitonov2025spinqubitperformanceerror}. Next, virtual qubits from the circuit ($q_i$) are mapped to physical qubits in the architecture, each identified by an index ($Q_0, Q_1, ..., Q_n$). Each quantum gate in the circuit must then be executed in a dedicated manipulation zone, indexed from 0 to the total number of qubits ($O_0, O_1, ..., O_n$), with each zone capable of handling two quantum states, enabling the execution of two-qubit gates. Therefore, qubits need to be moved through shuttling from their respective locations to a manipulation zone to interact. A shuttling operation from storage zone $i$ to manipulation zone $j$ is denoted as $\Omega(Q_i, O_j)$, which introduces a phase error $\delta C$ based on the shuttling velocity $v$ and the shuttling distance $L_s = d(Q_i, O_j)$ between the physical qubit and the manipulation zone. The resulting phase error is calculated using the error model given in Eq.~\eqref{eq:error_model}.


The mapping problem \cite{10.1145/3168822} consists of adding shuttling operations to move qubits from storage to manipulation zones for gate execution, ensuring both qubits are in the same zone for two-qubit gates, and returning them to storage afterwards—all facilitated by the shuttling bus system

The following subsections detail various circuit mapping techniques we are proposing, each progressively improving upon the previous, with the goal of minimizing the accumulated phase error $\delta C$ and the overall circuit execution time, which considers both the time required for shuttling operations and for executing the quantum gates in the circuit.

\subsection{Baseline}
This mapping strategy serves as the baseline for this work. The process iterates through each operation in the circuit and executes one gate at a time. For single-qubit gates applied to a virtual qubit stored in physical qubit $Q_i$, the qubit is moved to the nearest manipulation zone, $O_{k=i}$, using a shuttling operation $\Omega(Q_i, O_k)$ at a fixed velocity of 10 m/s \cite{langrock2023blueprint}. After the gate is executed, the qubit is returned to its original physical qubit position with a shuttling operation $\Omega(O_k, Q_i)$. For two-qubit gates involving qubits $Q_i$ and $Q_j$, both qubits are moved to the central manipulation zone $O_{k = \lceil (i+j)/2 \rceil}$ using shuttling operations $\Omega(Q_i, O_k)$ and $\Omega(Q_j, O_k)$, and then returned to their respective physical qubit locations at the same fixed velocity of 10 m/s.

\subsection{Parallel}
A major drawback of the \emph{Baseline Mapping} is that it executes gates sequentially, despite the architecture supporting parallel execution as long as the operations occur in different manipulation zones. To exploit this parallelism, the circuit is divided into slices, where each slice consists of gates that can be executed simultaneously. Qubits are moved to their nearest manipulation zone for single-qubit gates, just like in the baseline. For two-qubit gates involving qubits $Q_i$ and $Q_j$, they are moved to manipulation zone $O_{k = \text{max}(i,j)}$. This ensures the shuttling bus moves only in one direction, and every qubit in the slice is assigned to a manipulation zone that is located to the right of its current physical position, allowing all qubits to move simultaneously in parallel.

\subsection{Minimum Return}
In both the \emph{Baseline} and \emph{Parallel Mapping} strategies, qubits always return to their original physical qubit positions after executing each gate. Therefore, if the qubit stored in $Q_0$ were to interact with the rightmost qubit $Q_n$, we would need to shuttle across the whole shuttling bus twice (i.e. $\Omega(Q_0, O_n)$ and $\Omega(O_n, Q_0)$) adding a high error phase to the state.

The \emph{Minimum Return} strategy addresses this by dynamically reallocating qubits after gate execution. After determining the manipulation zones for each gate, as in \emph{Parallel Mapping}, the free physical qubits are sorted by their positions, and unallocated virtual qubits are assigned to the nearest free physical qubits. This assignment ensures all qubits move in a leftward direction after gate execution, minimizing unnecessary movement and enabling the parallel shuttling of qubits.

\subsection{Tunable Velocity}
All previous strategies use a fixed shuttling velocity of 10~m/s\cite{langrock2023blueprint}, but the phase error $\delta C$ depends on both the shuttling distance $L_s$ and velocity $v$. 

To this end, we employ the same movements as the \emph{Minimum Return}, but now, the shuttling velocity will depend on the maximum distance to shuttle. To do so, we compute all the distances to be shuttled, first for moving the qubits from the storage to the manipulation zone and after for returning them to the physical qubits (employing the minimum return policy). Then, by using the maximum shuttling distance and computing the derivative of the error model with respect to the velocity $\frac{\partial}{\partial v} \delta C$, we determine the optimal shuttling velocity.


\subsection{SWAP Return}
Lastly, to reduce the error added to each qubit and the total execution time even further, we modify the return operation (i.e., going from manipulation to storage zones) to take into account future interactions of the qubits.

For instance, if virtual qubits $q_i$ and $q_j$ interact in manipulation zone $O_k$, and the (sorted) physical qubits for their return are $Q_i'$ and $Q_j'$. The \emph{Minimum Return} strategy arbitrarily places $q_i$ into $Q_i'$ and $q_j$ into $Q_j'$, though it may not be optimal. For this, we compute the distances that future interactions of qubits $q_i$ and $q_j$ will require depending on the placement. Let $q_l$ in $Q_l$ be the next interaction for $q_i$, and $q_m$ in $Q_m$ be the next interaction for $q_j$. We assign $q_i$ to $Q_i'$ and $q_j$ to $Q_j'$ if $d(Q_i', Q_l) \leq d(Q_j', Q_m)$; otherwise, we assign $q_i$ to $Q_j'$ and $q_j$ to $Q_i'$, this reduces both the total phase error and the overall execution time by optimizing the return path based on upcoming interactions.

\subsection{Qubit Initial Placement}
We now address the issue of initial qubit placement, i.e., how to assign each virtual qubit to a physical qubit. As mentioned earlier, some mapping techniques (such as \emph{Minimum Return}, \emph{Tunable Velocity}, and \emph{Swap Return}) involve dynamic qubit placement, while others (like \emph{Baseline} and \emph{Parallel Mapping}) use a static placement. Regardless of the mapping strategy, a good initial placement can reduce both the phase error and the total execution time.

Given the 1D connectivity of the qubits, the initial placement can be framed as a Minimum Linear Arrangement problem \cite{harper1964optimal}, where the goal is to place the vertices of a graph on a line such that the sum of the distances between connected vertices is minimized. This problem is NP-hard \cite{lewis1983michael}, so we use heuristic approaches to find approximate solutions.

In our case, the input is the interaction graph of the quantum circuit \cite{Bandic2023}, where vertices represent qubits and edges represent interactions between them. The edges are weighted according to the urgency of the interaction, using an exponential decay function $2^{-l}$, where $l$ represents the layer in which the interaction occurs.

We propose to use the spectral layout method \cite{KOREN20051867} to compute the initial placement. The spectral layout is based on the eigenvectors of the Laplacian matrix $L=D-A$ (where $D$ is the degree matrix and $A$ is the adjacency matrix of the interaction graph). This method calculates the largest eigenvalues and corresponding eigenvectors, which are then used to assign coordinates to the graph’s vertices. Since we are working with a 1D qubit array, we use only the eigenvector associated with the first eigenvalue of the Laplacian matrix to determine the linear arrangement of the qubits.

\section{Results}

\begin{table}
    \caption{Parameters and ranges used in the experiments.}
    {\centering
    \begin{tabular}{c|c || c|c}
        \textbf{Architecture} &   & \textbf{Error Model} &  \\
        \textbf{Parameter} & \textbf{Value} & \textbf{Parameter} & \textbf{Value}\\
        \hline Distance qubit-qubit & 2 µm & $l_c^{\delta \omega}$ $^\dagger$ &  100 nm\\
        \hline Distance qubit-gate & 1 µm & $T_2^*$ $^\dagger$ & 20 µs\\
        \hline Shuttling velocity & 10 m/s & $L_{dot}$ $^\dagger$ & 20 nm\\
        \hline Single-qubit gate time & 20 ns & $E_{VS,0}$ $^\dagger$ & 100 µeV\\
        \hline Two-qubit gate time & 45 ns & $\bar d$ & 30 nm \\
        \hline \# Qubits & 10 -- 30 & $a_x$ $^\dagger$ & 0.05 $\pi$/nm
    \end{tabular}}
    
    {\footnotesize $^\dagger$ These parameters are taken from \cite{langrock2023blueprint}.}
    \vspace{-15pt}
    \label{tab:parameters}
\end{table}

\begin{figure*}
    \centering
    \includegraphics[width=\linewidth]{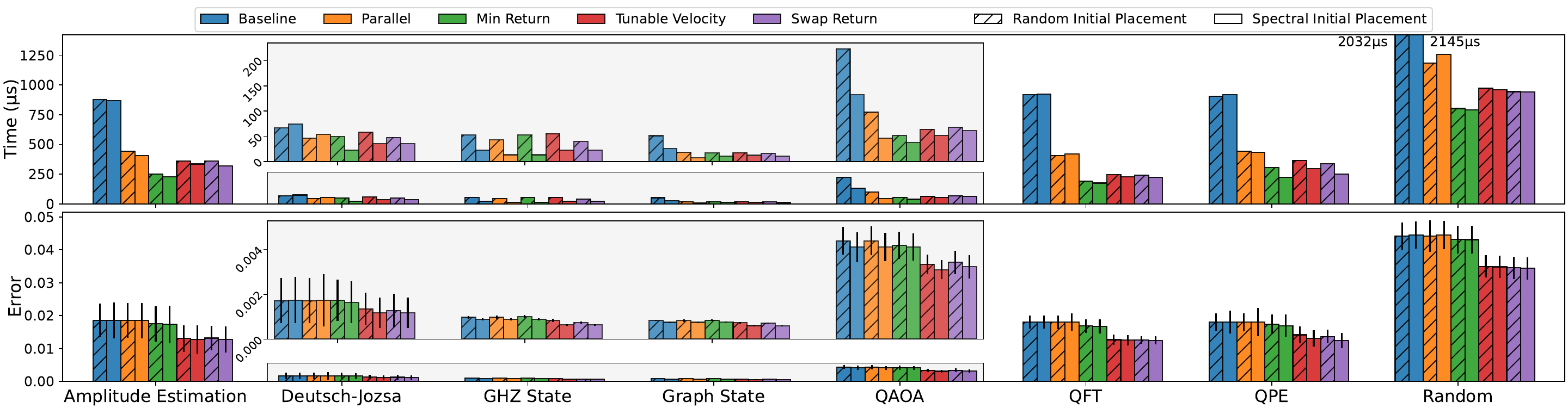}
    \vspace{-16pt}
    \caption{Performance of the proposed mapping strategies several benchmarks compiled into a 16 qubits architecture.}
    \label{fig:mapping_q16}
    \vspace{-20pt}
\end{figure*}

In this section, we present the results of our experiments aimed at evaluating the effectiveness of the proposed mapping strategies and the importance of initial qubit placement in quantum circuits. We selected seven quantum algorithms from the MQT Bench \cite{quetschlich2023mqtbench} as benchmarks to assess the performance of the mapping techniques. 
The architectural parameters used in the experiments, such as the number of physical qubits, manipulation zones, and the distance between qubits, are aligned with the size of the circuits being mapped (i.e. a circuit with 16 qubits will correspond to an architecture with 16 qubits and 16 manipulation zones). 
A summary of all the simulation parameters is provided in Table \ref{tab:parameters}. 

\subsection{Performance Comparison of Mapping Strategies}
In the first experiment, we fixed the circuit size to 16 qubits and compiled the selected benchmarks using each proposed mapping strategy. We measured two key metrics: the total execution time (including both the time for shuttling and gate execution) and the phase error introduced to the qubits, reporting both the mean and standard deviation. Figure \ref{fig:mapping_q16} presents the total time (upper row) and phase error (lower row) for each circuit, comparing the performance with a random initial placement (averaged over 10 runs) and the proposed spectral initial placement.

The results indicate that the \emph{Minimum Return} strategy achieved the best performance in terms of execution time ($2.92 \times$ over baseline). However, when considering the phase error, the \emph{Tunable Velocity} ($1.28 \times$) and \emph{Swap Return} ($1.32 \times$) strategies showed significantly lower errors, suggesting that the optimal shuttling velocity is generally below 10 m/s for circuits of this size. This illustrates that while minimizing execution time is important, an aggressive reduction in time by increasing shuttling speed can lead to higher phase errors. Notably, the \emph{Swap Return} strategy outperformed \emph{Tunable Velocity} in most cases, although there are instances where this is reversed. The effect of initial placement is less conclusive: while the spectral initial placement typically outperforms random placement, the results vary across benchmarks.

\subsection{Impact of Initial Qubit Placement}
\begin{figure}
    \centering
    \includegraphics[width=\linewidth]{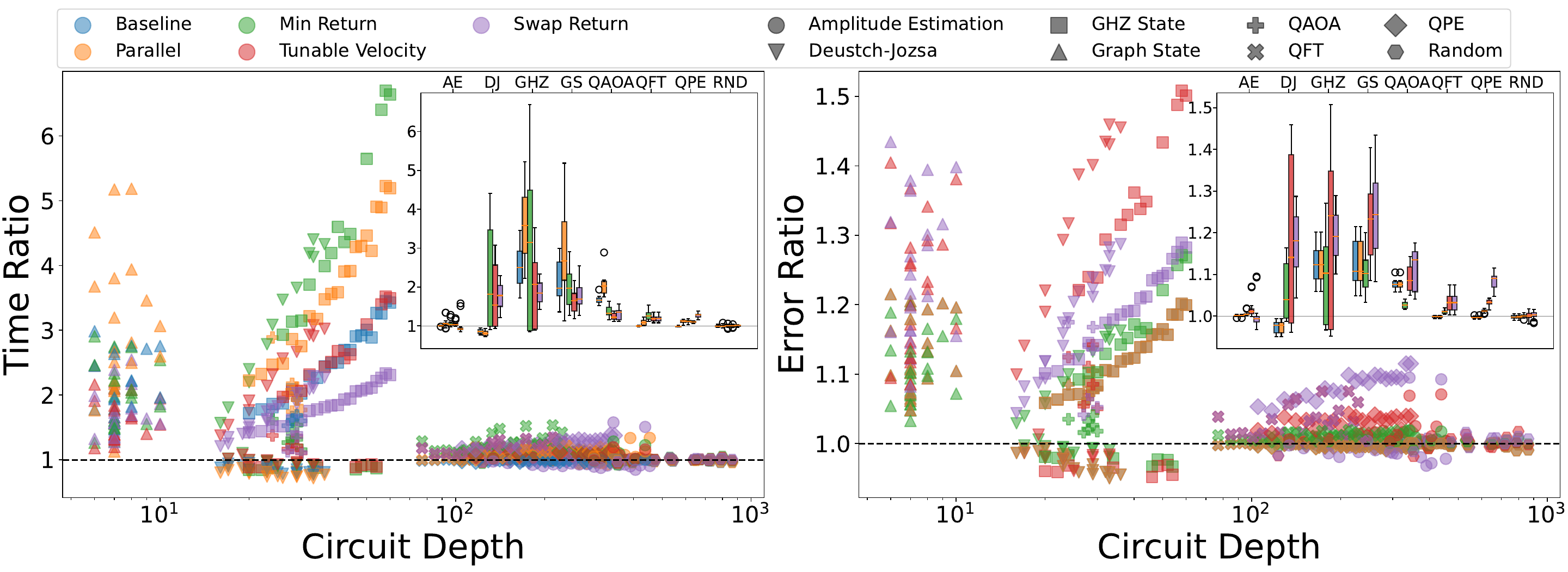}
    \vspace{-16pt}
    \caption{Initial placement impact for an increasing number of qubits.}
    \label{fig:initial_placement_experiment}
    \vspace{-15pt}
\end{figure}

In the second experiment, we focused on the impact of the initial placement across an increasing number of qubits, ranging from 10 to 30. Figure \ref{fig:initial_placement_experiment} illustrates the improvement ratio (random performance/spectral performance), showing how often the spectral initial placement outperforms random placement. The x-axis represents the depth of the circuit, while different markers correspond to specific benchmarks, and the colors indicate the mapping strategy employed.

Several clusters emerge in the results. The Graph State circuit, which has the shortest depth, consistently shows a great improvement with spectral initial placement, outperforming random placement in both execution time and phase error across all mapping strategies. A second cluster includes mid-sized circuits, such as Deutsch-Jozsa, QAOA, and GHZ, where significant improvements are observed for certain strategies, including \emph{Swap Return}. The final cluster consists of larger circuits, including QPE, QFT, and Random Circuits, where the impact of initial placement is less pronounced. These findings highlight that the effectiveness of initial placement strategies depends on the depth and complexity of the quantum circuit.

\section{Conclusions}
In this work, we introduced and evaluated several mapping strategies to optimize qubit shuttling for silicon spin qubits in scalable quantum architectures. These strategies aimed to minimize phase errors while reducing the total execution time of quantum circuits. Among the proposed mapping techniques, the \emph{Swap Return} proved to be the most robust, achieving an optimal balance between minimizing shuttling-induced phase errors and maintaining competitive execution times. 

Finally, the importance of initial qubit placement was explored, showing that a well-chosen initial placement can enhance the performance of mapping strategies. The spectral layout for the initial placement consistently outperformed random placement in most cases, particularly for shorter circuits.

\bibliographystyle{IEEEtran}
\bibliography{IEEEabrv, references}

\end{document}